\documentclass[cits]{PoS}
\usepackage[T1]{fontenc}
\usepackage{cite}
\usepackage{listings}

\newcommand{\Intel}{{Intel\textsuperscript{\textregistered}\ }}
\newcommand{\XeonPhi}{{Xeon Phi{\texttrademark}\ }}

\title{A performance evaluation of CCS QCD Benchmark on the COMA (\Intel \XeonPhi, KNC) system}

\ShortTitle{A performance evaluation of CCS QCD Benchmark on the COMA system}

\author{
Taisuke Boku$^{a,b}$,
\speaker{Ken-Ichi Ishikawa}$^{,c,d}$,
Yoshinobu Kuramashi$^{a,e}$,
Lawrence Meadows$^{f}$,
Michael D`Mello$^{f}$,
Maurice Troute$^{f}$,
Ravi Vemuri$^{f}$\\
{$^{a}$}Graduate School of Systems and Information Engineering, University of Tsukuba, Tsukuba, Ibaraki 305-8573, Japan\\
{$^{b}$}Center for Computational Sciences, University of Tsukuba, Tsukuba, Ibaraki 305-8577, Japan\\
{$^{c}$}Graduate School of Science, Hiroshima University, Higashi-Hiroshima, Hiroshima 739-8526, Japan\\
{$^{d}$}Core of Research for the Energetic Universe, Hiroshima University, Higashi-Hiroshima, Hiroshima 739-8526, Japan\\
{$^{e}$}Faculty of Pure and Applied Sciences, University of Tsukuba, Tsukuba, Ibaraki 305-8571, Japan\\
{$^{f}$}Intel Corporation, USA\\
E-mail: \email{ishikawa@theo.phys.sci.hiroshima-u.ac.jp}}

\abstract{
 {\normalsize        % 
 \vspace*{-29.5em}   % 
 \begin{flushright}  % This part must be removed for PoS submission
 HUPD-1613, UTHEP-699, UTCCS-P-95
 \end{flushright}    %
 \vspace*{29em}\     %
 }
The most computationally demanding part of Lattice QCD simulations is solving quark propagators. 
Quark propagators are typically obtained with a linear equation solver utilizing HPC machines. 
The CCS QCD Benchmark is a benchmark program solving the Wilson-Clover quark propagator, 
and is developed at the Center for Computational Sciences (CCS), University of Tsukuba. 
We optimized the benchmark program for a \Intel \XeonPhi (Knights Corner, KNC) system 
named ``COMA (PACS-IX)'' at CCS Tsukuba under the Intel Parallel Computing Center program. 
A single precision BiCGStab solver with the overlapped Restricted Additive Schwarz (RAS) 
preconditioner was implemented using SIMD intrinsics, OpenMP and MPI in the offload mode. 
With the reverse-offloading technique, we could reduce the communication and offloading overheads.
We observed a performance of $\sim 200$ GFlops sustained for the Wilson-Clover hopping 
matrix multiplication on the lattice sizes larger than $24^3\times 32$ on a sinlge card of 
the COMA system.
A good weak scaling perofmace was observed on the local lattice sizes larger 
than $24^3\times 32$.
}

\FullConference{34th annual International Symposium on Lattice Field Theory\\
		24-30 July 2016\\
		University of Southampton, UK}

\begin{document}

\section{Introduction}
The success of Lattice QCD simulations owes much to the development of 
numerical algorithms and optimization for the quark solver, and evolution of 
HPC machines.  
We have developed a quark solver benchmark program called ``CCS QCD Benchmark'' (CCS-QCD)~\cite{CCS:QCD}, 
which solves the Wilson-Clover quark propagator, 
at the Center for Computational Sciences (CCS), University of Tsukuba. 
This is designed to be as simple as possible and is written in plain Fortran 90 
so that new algorithms or new HPC architectures can be evaluated quickly with 
this benchmark program. 

A new architecture system equipped with the \Intel \XeonPhi (Knights Corner, KNC) co-processor 
cards has been installed at CCS in 2014. The name of the system is ``COMA (PACS-IX)''~\cite{CCS:COMA}.
This is the ninth system of the PACS/PAX series~\cite{PACSSERIES}.
The \Intel \XeonPhi (Knights Corner, KNC) co-processor cards is based on 
the Intel Many Integrated Core\texttrademark (MIC) architecture, 
and has many physical cores compatible to x86-64  on a chip.
Although the programming model is common to x86-64 based systems, it requires some tuning 
tips to fully extract the many core performance. 
The communication among the co-processor cards requires HOST-HOST and HOST-KNC communication 
like GPGPU computing. 
There have been a lot of studies on the QCD program for KNC systems in 
the past few years~\cite{QCD:KNC:JOO,QCD:KNC}.
We optimize the CCS-QCD program for the COMA system to extract the best performance. 
The basic strategy to optimize the CCS-QCD program for the KNC system, such as
prefetching, threading, SIMD-vectorization \textit{etc.}, is almost 
the same as those studied in 
Refs.~\cite{QCD:KNC:JOO,QCD:KNC}.
This year a first result
using the next generation \Intel \XeonPhi system (Knights Landing, KNL) 
has been presented in this conference~\cite{DeTar:2016ndn}.
In this talk, we especially focus on the parallel performance of the CCS-QCD on the COMA system. 

There are three running modes for a typical KNC system in executing a MPI program; 
``native'',  ``symmetric'', and ``offload modes''.
We employ the ``{offload mode}'' for the CCS-QCD to utilize the single precision
acceleration to the solver algorithm, where the single precision solver is added 
and involved as the preconditioner to the double precision solver.
The directive based programming is available as 
the Language Extensions for Offload (LEO) in the \Intel compiler.
The total amount of the code modification on the original code is minimized 
by offloading the single precision solver to the accelerator.
The performance of the whole program relies on the performance of the single precision 
solver added. 

To have the best performance, together with the tuning and the optimization for 
the computational part on the co-processor, 
the MPI communication among the co-processor cards in the offload mode must be 
considered as the MPI functions cannot be used in offload regions.
Typically the MPI functions and manipulating data are 
handled by the host CPU code in the offload mode.
The data transfer from a host CPU to the co-processor on the host and vice versa 
can be done only at the beginning or end of the offload region using the directive.
The MPI communication splits the entire solver code into many parts of the offload regions.
The offloading overhead could be a bottleneck of the performance.

To get rid of the limitation in the offload mode and reduce the offloading overhead, 
we implement a proxy server code running on the host CPU which handles the request of
the MPI-communication from the offload region. 
The communication between the host proxy and the offloaded code on the KNC 
is done via the Symmetric Communications InterFace (SCIF)~\cite{INTEL:SCIF}. 
With the proxy code, the entire code of the single precision solver can be packed 
in a single offload region and the MPI requests are reversely offloaded to 
the host proxy. This strategy is called reverse-offloading. 
The proxy and reverse-offloding have been introduced in Ref.~\cite{QCD:KNC:JOO}.
This strategy also enables us to apply the communication-computation overlapping.
The single precision solver is programmed in this way.
The tuned code for the COMA system is also available at~\cite{CCS:QCD}.

This paper is organized as follows. 
In the next section, we  mainly describe the details of the reverse-offloading and 
the communication-computation overlapping.
In section~\ref{sec:Results} we show the performance of the code and 
summarize this paper.

\section{Tuning the CCS-QCD for the COMA system}
\label{sec:Tuning}
The COMA system is composed of 393 computational nodes equipped with
two CPUs (\Intel Xeon E5-2680v2) and
two \XeonPhi 7110P co-processor (KNC) cards. 
All the nodes are connected by full-bisection bandwidth of Fat-Tree network of InfiniBand FDR.
The theoretical peak performance is 1.001 PFlops including 157.2 TFlops of CPUs.
Making a full use of the co-processors (84\% of the system peak) is inevitable
to get the best performance of the system.

The CCS QCD Benchmark (CCS-QCD) implements the BiCGStab solver algorithm for 
the even-odd site preconditioned Wilson-Clover quark matrix in double precision. 
The code is written in Fortran 90 Language and parallelized in 
the $X$, $Y$, and $Z$ directions using MPI. 
We replace the double precision BiCGStab solver algorithm to 
the double precision flexible BiCGStab solver algorithm~\cite{FBICGSTAB}.
We add a single precision BiCGStab solver, to be offloaded to the co-processor, 
as the preconditioner to the flexible BiCGStab. 
Hereafter, we refer to the single precision BiCGStab solver as the solver for simplicity,
and we focus on the solver performance.

The single precision solver is further preconditioned with the even-odd site and 
the overlapped Restricted Additive Schwarz (RAS) preconditioning~\cite{Osaki:2010vj}. 
The solver is written in the C/C++ language to make use of the SIMD intrinsic functions 
of the Intel C/C++ compiler. The SIMD length is 16 for float and we embed four time slices 
of the spinor and gauge link fields into a SIMD vector (\texttt{\_\_m512});
four time slices of a two-component spinor at a color index, and
four time slices of first two column elements of a SU(3) matrix at a row-color index. 
The so-called SU(3)-reconstruction technique is employed.
The many cores of the \XeonPhi are organized by OpenMP parallel threading. 
Loop blocking on the lattice sites and explicit prefetching is employed to enhance 
the use of the cache locality in a physical core.
We use one co-processor within a MPI process. 
Thus two MPI processes are located on the COMA node.
The co-processor identifier (0 or 1) is assigned to the MPI process according to the 
even/odd-ness of the MPI RANK.

\begin{figure}[t]
    \centering
\vspace*{-2em}
{
\lstset{language=C,basicstyle=\fontsize{6}{3.8}\selectfont\ttfamily\bfseries,numbers=left}
\begin{lstlisting}[frame=single]
extern ``C'' void
assign_inv_mult_offl_ras_solver( const float      *kappa,
                                 const float      *stol,
                                       int        *iter,
                                 const mic_wqf_eo *swe,
                                       mic_wqf_eo *sve)
{
    int offload_signal = 0xF;
    const float skappa1 = *kappa;
    const float stol1 = *stol;
    int iter1 = *iter;

    ////////////////////////////////////////////////////////////////////////
    // Asynchronous offloading of native code
    //   proxy server is running parallel to the native code.
    ////////////////////////////////////////////////////////////////////////
    static mic_wqf_eo *swe1, *sve1;
    if (swe1 == 0)
    {
    #pragma offload target(mic:mic_targetid) \
      nocopy(swe1[0:1] : alloc_if(1) free_if(0) preallocated targetptr) \
      nocopy(sve1[0:1] : alloc_if(1) free_if(0) preallocated targetptr)
    {
      swe1 = (mic_wqf_eo *)_mm_malloc(sizeof(mic_wqf_eo), 64);
      sve1 = (mic_wqf_eo *)_mm_malloc(sizeof(mic_wqf_eo), 64);
    }
    }
      
    #pragma offload target(mic:mic_targetid) \
        nocopy(scif_mic) \
        in( swe [0:1] : into(swe1[0:1]) alloc_if(0) free_if(0) targetptr) \
        out(sve1[0:1] : into(sve [0:1]) alloc_if(0) free_if(0) targetptr) \
        in(skappa1,stol1) \
        inout(iter1) \
        signal(offload_signal)
    {

        assign_inv_mult_eoprec_wd_bicgstab_mic(&skappa1,&stol1,&iter1,sve1,swe1);

    }

    ////////////////////////////////////////////////
    // process proxy 
    // receive requests from the native code located
    // in the above offload region.
    ////////////////////////////////////////////////
    process_cmds();

    ////////////////////////////////////////////////
    // wait for finishing of offloading
    ////////////////////////////////////////////////
    #pragma offload_wait target(mic:mic_targetid) wait(offload_signal)
....
\end{lstlisting}
}
\vspace*{-1em}
\caption{Offloading part in the host CPU code.}
\label{fig:OFFLOADCODE}
\end{figure}

\subsection{Reverse-offloading}
The reverse offloading~\cite{QCD:KNC:JOO} is implemented as follows. 
The solver code to be offloaded is essentially the same as that in the native mode except for 
the MPI communication part. 
Instead of calling the MPI APIs, the solver code calls wrapper functions similar to the MPI, 
in which the communication requests from the KNC are translated into the MPI-requests,
and they are reversely offloaded to the proxy on the host CPU via SCIF. 
We refer to this communication API as SCIF simply.

\begin{figure}[t]
    \centering
\vspace*{-2em}
    \includegraphics[scale=0.45]{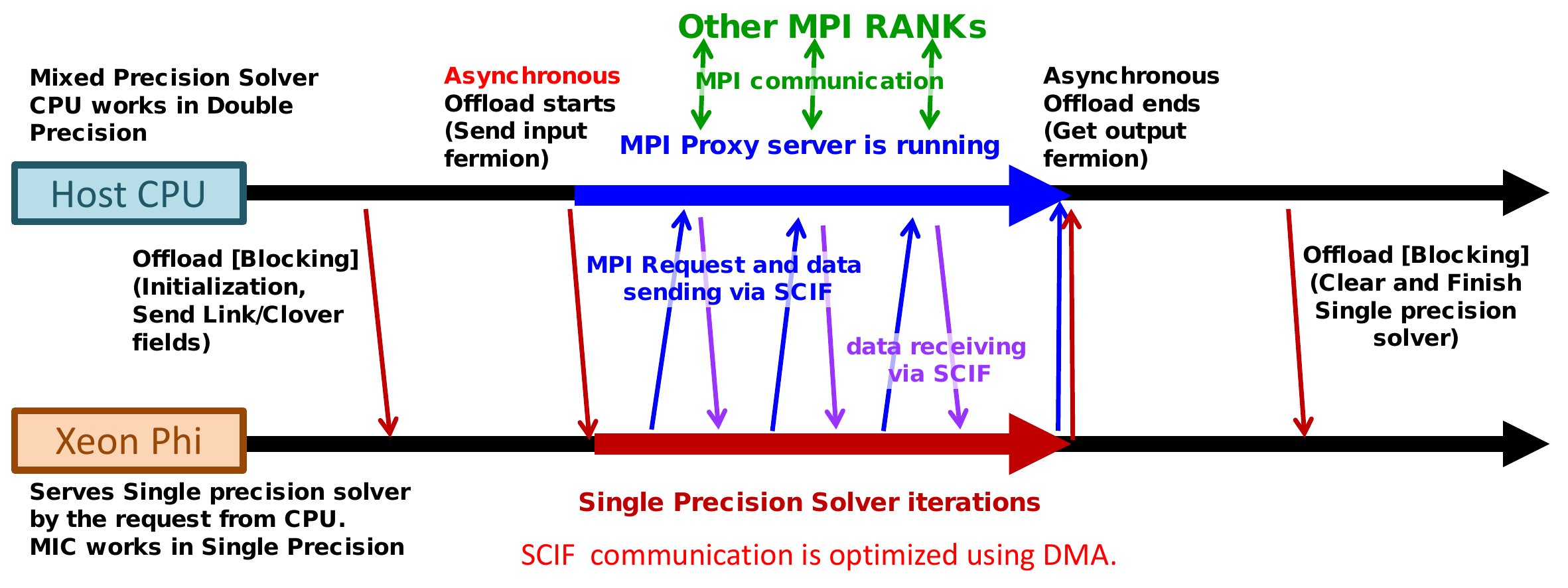}
\vspace*{-1em}
    \caption{Reverse offloading in the solver.}
    \label{fig:revoffloading}
\end{figure}

To launch the solver to the co-processor and the proxy server on the host CPU simultaneously, 
asynchronous offloading is used. 
Figure~\ref{fig:OFFLOADCODE} shows the code snippet launching 
the single precision solver and run the proxy asynchronously, where ``\texttt{\#pragma offload*}'' 
are the directive for offloading.
The schematic diagram of the program behavior is shown in Figure~\ref{fig:revoffloading}.

The input spinor vector memory is allocated on the co-processor in the lines 19--27, where
the pointers \texttt{swe1} and \texttt{sve1} are kept on the co-processor across the offload region.
In the lines 29--40, the single precision solver is called on the co-processor after 
the data transmission (copy in) of the source vector 
to \texttt{swe1} of the co-processor memory from \texttt{swe} of the host memory 
(the start point of the blue and red horizontal arrows in Figure~\ref{fig:revoffloading}).
This offload region is asynchronous as indicated by the directive \texttt{signal(offload\_signal)} and 
the control is non-blocking over the host CPU. 
Then,
without waiting for the termination of the offload region, the host CPU calls \texttt{process\_cmds} at the line 47, 
which is the proxy function, where the MPI requests from the solver \texttt{assign\_inv\_mult\_...\_mic} 
located at the line 38 are processed (the blue and red horizontal arrows and the communication arrows between them in Figure~\ref{fig:revoffloading}).
When the solver converges, the solver sends a termination signal to the proxy \texttt{process\_cmds}. 
After receiving the termination command in the proxy, the control of CPU moves 
to the line 52, where CPU waits for the termination signal from the offload region asynchronously 
emitted in the lines 29--40. This includes the completion of the data transmission (copy out) of 
the solution vector to \texttt{sve} of the host memory from \texttt{sve1} of the co-processor memory
(the end point of the blue and red horizontal arrows in Figure~\ref{fig:revoffloading}).

\subsection{Communication-computation overlapping}
Using the reverse-offloading and the SCIF, the co-processor can use the communication 
functions similar to the non-blocking MPI and DMA transmission functions. 
With this function, we implement the communication-computation overlapping 
in the Wilson-Clover Dirac hopping matrix multiplication (MULT).

\begin{figure}[t]
    \centering
\vspace*{-0.0em}
    \includegraphics[scale=0.45]{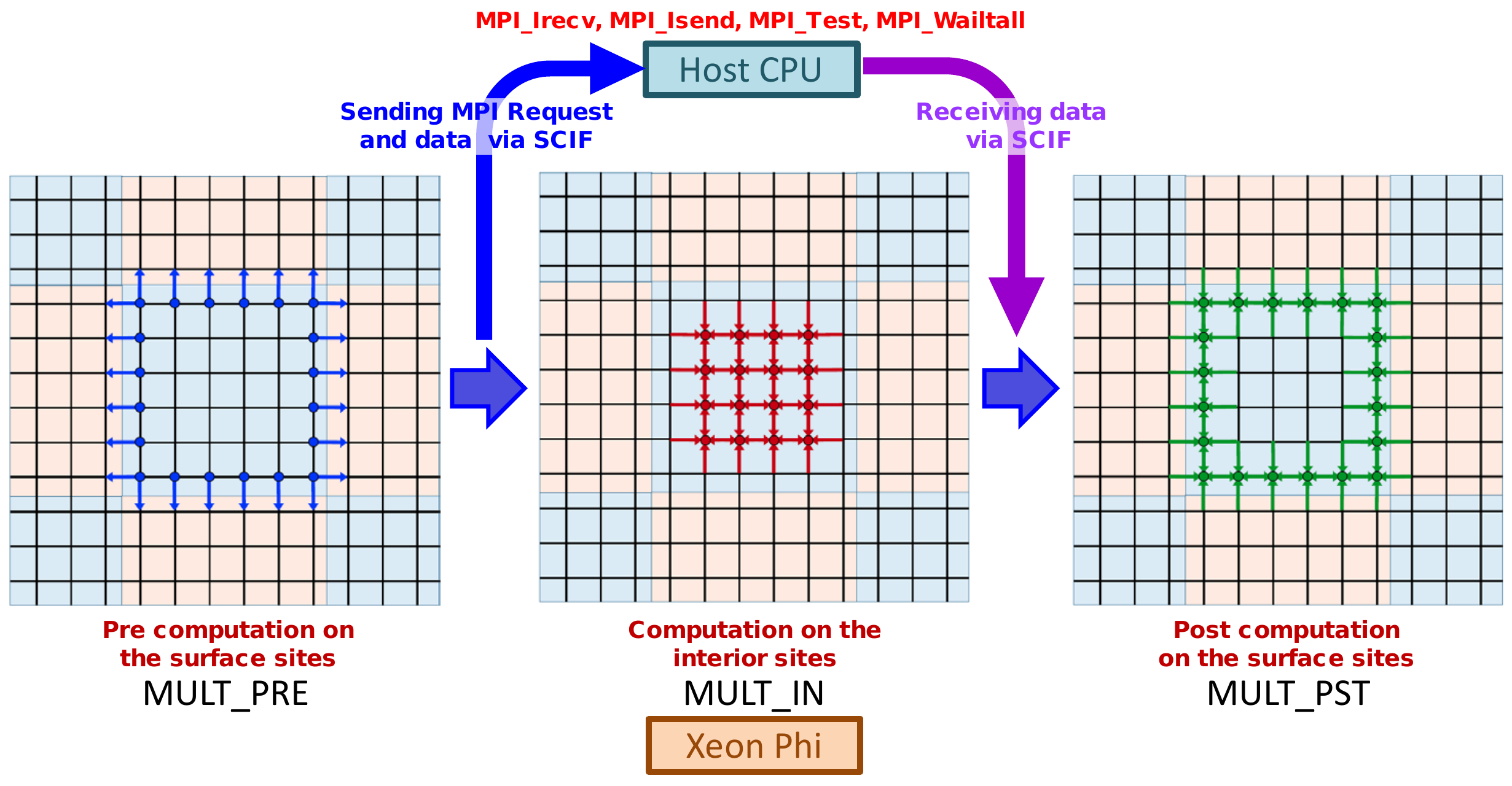}
\vspace*{-1em}
    \caption{Task separation in the hopping matrix multiplication.}
    \label{fig:MultSplit}
\end{figure}

The implementation of the communication-computation overlapping for the 
the Wilson-Clover Dirac hopping matrix is usually organized by splitting the task into two pieces;
(i) the computation within the node, and (ii) the computation using the data from other nodes. 
In this case the conditional branch statements are required to distinguish the site location 
and the direction of the hopping operation. 
To avoid the use of the conditional branch, as the KNC has a rather less performance on 
the conditional branch,
we split the task in slightly different manner as depicted in 
Figure~\ref{fig:MultSplit}. The lattice sites in a process are split into the interior sites 
and the surface sites. 
In the ``\texttt{MULT\_PRE}'' the spin-projection and data packing to the send buffer 
are carried out.
After submitting the MPI request to SCIF in no-blocking way, 
the hopping computation in the interior sites continues in ``\texttt{MULT\_IN}''. 
In this region there is no conditional branches on the hopping directions.
The communication request is processed on the host CPU during the computation.
After receiving the data from the host CPU, the stencil computation on the surface sites
follows in ``\texttt{MULT\_PST}''. 
With the task separation a good performance is expected in ``\texttt{MULT\_IN}''.

\section{Results and Summary}
\label{sec:Results}

We benchmark the tuned code on the COMA system. 
The performance of the hopping multiplication (MULT) is measured
varying the local lattice size and the parallelism.

\begin{figure}[t]
    \centering
\vspace*{-2em}
    \includegraphics[scale=0.52]{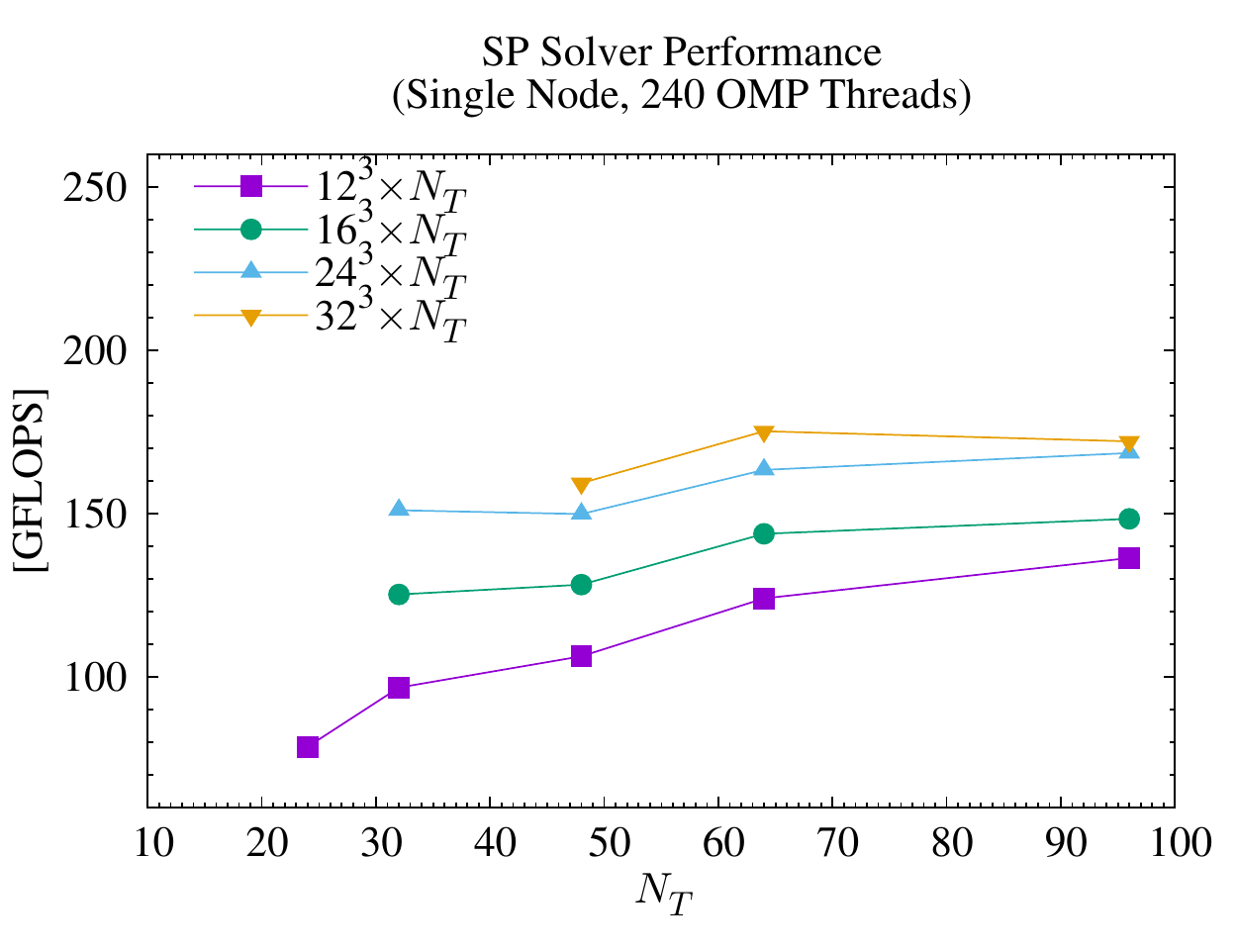}
\hfill
    \includegraphics[scale=0.52]{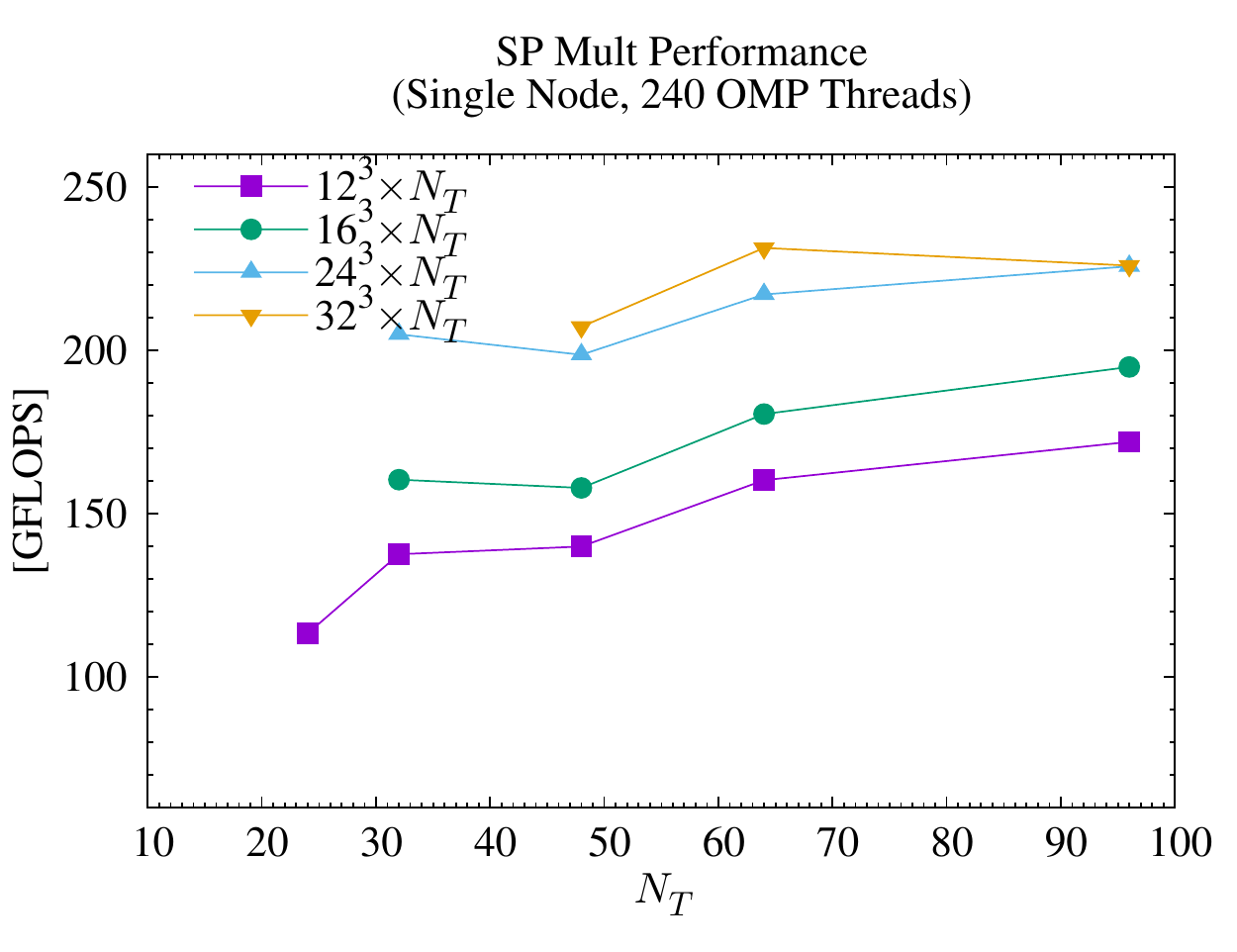}
\vspace*{-1em}
    \caption{Solver (left) and MULT (right) performance with single process.}
    \label{fig:SingleProcPerf}
\end{figure}

The single node performance, which is the baseline for the parallel computation, is shown in 
Figure~\ref{fig:SingleProcPerf}. In this case there is no task splitting in 
the MULT computation and periodic boundary condition is imposed.
Basic optimizations used for KNC systems are applied and we obtain $\sim 200$ GFlops 
for sufficiently larger lattice sizes ($> 40000$ sites) in a process.

The parallel performance is tested with the weak-scaling benchmarking. 
We test two local lattice sizes for the spatial size $N_S=12$ and 24, and vary the temporal local lattice size.
The parallelisms tested are
$1\times 1\times 1$ (baseline),
$2\times 1\times 1$,
$2\times 2\times 1$,
$2\times 2\times 2$,
$4\times 2\times 2$, and
$4\times 4\times 2$ for the weak-scaling.

\begin{figure}[t]
    \centering
\vspace*{-0.5em}
    \includegraphics[scale=0.52]{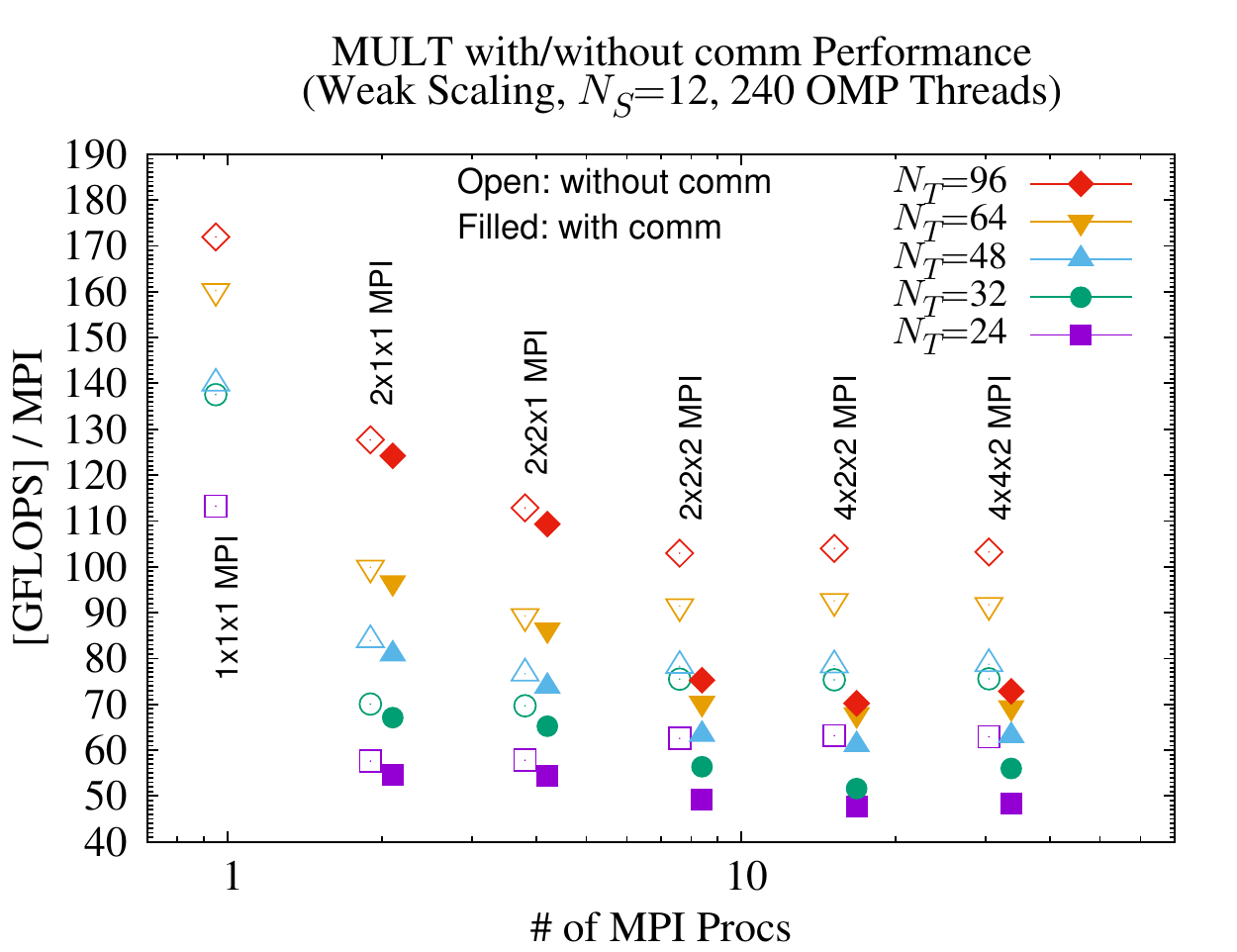}
\hfill
    \includegraphics[scale=0.52]{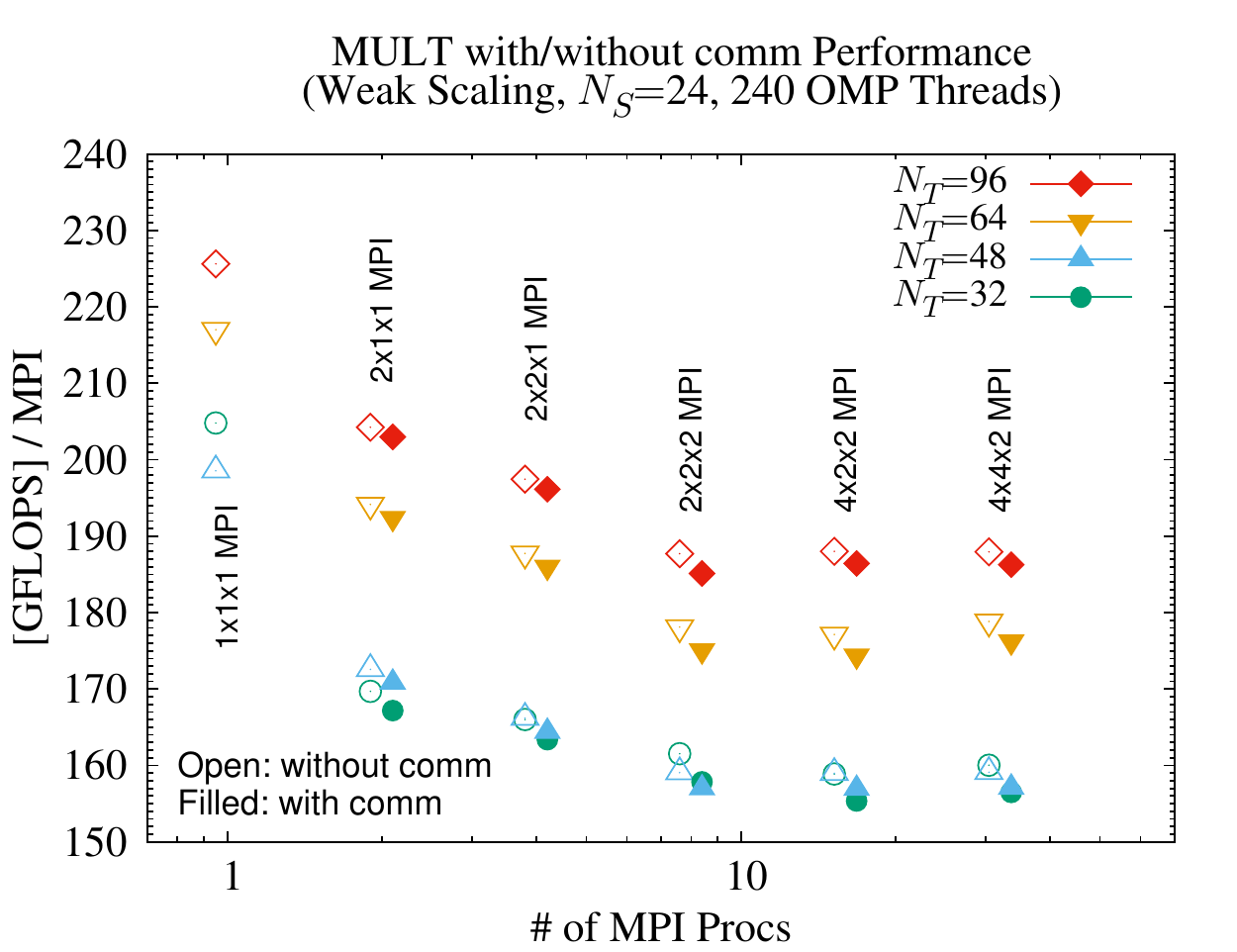}
\vspace*{-1em}
    \caption{Performance of MULT in weak-scaling test ($N_S=12$ (left) and $N_S=24$ (right)).}
    \label{fig:WeakScalingPerf}
\end{figure}

Figure~\ref{fig:WeakScalingPerf} shows the performance of MULT. 
Filled symbols include the communication time of waiting for receiving data before \texttt{MULT\_PRE},
open ones are the timing without the communication time. 
For the results with small local volume of $N_S=12$ (left panel), a large gap between 
the open and filled symbols appears for the cases with three-dimensional parallelism.
On the other hand, the performance degradation (open vs filled) is small for sufficiently 
larger local volume (right panel).
This indicates the communication time is hidden behind the computation time. 
We need $N_S=24$ for the communication hiding.
We observe $\sim 190$ GFlops for the MULT performance at the local lattice size $24^3\times 96$.
The weak-scaling behavior is good as seen from the cases with three-dimensional parallel partitioning.

We have implemented the single precision solver code for \Intel \XeonPhi (KNC) 
and applied it into the CCS QCD Benchmark. 
By using the reverse-offloading technique and the SCIF interface, we have achieved $\sim 190$ GFlops
for the Wilson-Clover hopping multiplication with a $24^3\times 96$ local lattice size on the COMA system.
The next generation \Intel \XeonPhi (KNL) system named ``Oakforest-PACS''
will appear soon in the Joint Center for Advanced High Performance Computing (JCAHPC) in Japan~\cite{KNL:OFP}.
It could be interesting to optimize the CCS QCD Benchmark to the ``Oakforest-PACS'' system.

\vspace*{-1em}
\acknowledgments
\vspace*{-1em}
This work was supported by the Intel Parallel Computing Center 
at the Center for Computational Sciences (CCS), University of Tsukuba~\cite{ICCPCCS}.


\begin{thebibliography}{99}

\bibitem{CCS:QCD}
CCS QCD Benchmark program, Center for Computational Sciences (CCS), University of Tsukuba, 
\href{https://www.ccs.tsukuba.ac.jp/eng/research-activities/published-codes/qcd/}
{https://www.ccs.tsukuba.ac.jp/eng/research-activities/published-codes/qcd/}.

\bibitem{CCS:COMA}
COMA System, Center for Computational Sciences (CCS), University of Tsukuba, 
\href{http://www.ccs.tsukuba.ac.jp/eng/research-activities/supercomputers/}
{http://www.ccs.tsukuba.ac.jp/eng/research-activities/supercomputers/}.

\bibitem{PACSSERIES}
History of PACS/PAX Series, Center for Computational Sciences (CCS), University of Tsukuba, 
\href{http://www.ccs.tsukuba.ac.jp/eng/research-activities/projects/ha-pacs/history/}
{http://www.ccs.tsukuba.ac.jp/eng/research-activities/projects/ha-pacs/history/}.

%\bibitem{Intel:MICarch}
%\Intel  Many Integrated Core Architecture, Intel Corporation,
%\href{http://www.intel.com/content/www/us/en/architecture-and-technology/many-integrated-core/intel-many-integrated-core-architecture.html}
%{http://www.intel.com/content/www/us/en/architecture-and-technology/many-integrated-core/intel-many-integrated-core-architecture.html}

%%%%%%%%%%%%%%%%%%%%%%%%%%%%%%%%%%%%%%%%%%%%%
% QCD on KNC 
%%%%%%%%%%%%%%%%%%%%%%%%%%%%%%%%%%%%%%%%%%%%%
\bibitem{QCD:KNC:JOO}
B.~Jo\'{o} {\it et al.}, 
%D.~D. Kalamkar, K.~Vaidyanathan, M.~Smelyanskiy, K.~Pamnany, V.~W.
%  Lee, P.~Dubey, and I.~Watson, William, 
%``{Lattice QCD on Intel Xeon Phi Coprocessors},'' 
  in \emph{Supercomputing}, ser. Lecture Notes in Computer
  Science, J.~M. Kunkel, T.~Ludwig, and H.~W. Meuer, Eds.,
  Springer Berlin Heidelberg, 2013, vol. 7905, pp. 40--54,
  \href{http://dx.doi.org/10.1007/978-3-642-38750-0_4}{http://dx.doi.org/10.1007/978-3-642-38750-0\_4}.

\bibitem{QCD:KNC}
  S.~Heybrock, B.~Jo\'{o}, D.~D.~Kalamkar, M.~Smelyanskiy, K.~Vaidyanathan, T.~Wettig and P.~Dubey,
  %``Lattice QCD with Domain Decomposition on Intel Xeon Phi Co-Processors,''
  {doi:10.1109/SC.2014.11}
  [arXiv:1412.2629 [hep-lat]];
  R.~Li and S.~Gottlieb,
  %``Staggered Dslash Performance on Intel Xeon Phi Architecture,''
  PoS LATTICE {\bf 2014} (2015) 034
  [arXiv:1411.2087 [hep-lat]];
  H.~Jeong {\it et al.},
  %``Performance of Kepler GTX Titan GPUs and Xeon Phi System,''
  PoS LATTICE {\bf 2013} (2014) 423
  [arXiv:1311.0590 [physics.comp-ph]];
  D.~Barthou {\it et al.},
  %``Automated Code Generation for Lattice Quantum Chromodynamics and beyond,''
  J.\ Phys.\ Conf.\ Ser.\  {\bf 510} (2014) 012005
  [arXiv:1401.2039 [hep-lat]];
  S.~Mukherjee, O.~Kaczmarek, C.~Schmidt, P.~Steinbrecher and M.~Wagner,
  %``HISQ inverter on Intel$\scriptsize{^{\circledR}}$ Xeon Phi$\scriptsize{^{TM}}$ and NVIDIA$\scriptsize{^{\circledR}}$ GPUs,''
  PoS LATTICE {\bf 2014} (2015) 044
  [arXiv:1409.1510 [cs.DC]];
  O.~Kaczmarek, C.~Schmidt, P.~Steinbrecher and M.~Wagner,
  %``Conjugate gradient solvers on Intel Xeon Phi and NVIDIA GPUs,''
  arXiv:1411.4439 [physics.comp-ph];
  Y.~C.~Jang {\it et al.} [SWME Collaboration],
  %``Code Optimization on Kepler GPUs and Xeon Phi,''
  PoS LATTICE {\bf 2014} (2014) 035
  [arXiv:1411.2223 [hep-lat]];
  P.~Arts {\it et al.},
  %``QPACE 2 and Domain Decomposition on the Intel Xeon Phi,''
  PoS LATTICE {\bf 2014} (2015) 021
  [arXiv:1502.04025 [cs.DC]];
  P.~Boyle, A.~Yamaguchi, G.~Cossu and A.~Portelli,
  %``Grid: A next generation data parallel C++ QCD library,''
  PoS LATTICE {\bf 2015} (2015) 023
  [arXiv:1512.03487 [hep-lat]];
  M.~Schr\"{o}ck, S.~Simula and A.~Strelchenko,
  %``Accelerating Twisted Mass LQCD with QPhiX,''
  PoS LATTICE {\bf 2015} (2016) 030
  [arXiv:1510.08879 [hep-lat]];
  S.~Heybrock, M.~Rottmann, P.~Georg and T.~Wettig,
  %``Adaptive algebraic multigrid on SIMD architectures,''
  PoS LATTICE {\bf 2015} (2016) 036
  [arXiv:1512.04506 [physics.comp-ph]];
  H.~Kobayashi, Y.~Nakamura, S.~Takeda and Y.~Kuramashi,
  %``Optimization of Lattice QCD with CG and multi-shift CG on Intel Xeon Phi Coprocessor,''
  PoS LATTICE {\bf 2015} (2016) 029.

%%%%%%%%%%%%%%%%%%%%
% QCD on KNL
%%%%%%%%%%%%%%%%%%%%
%\cite{DeTar:2016ndn}
\bibitem{DeTar:2016ndn}
  C.~DeTar, D.~Doerfler, S.~Gottlieb, A.~Jha, D.~Kalamkar, R.~Li and D.~Toussaint,
  %``MILC staggered conjugate gradient performance on Intel KNL,''
  arXiv:1611.00728 [hep-lat].
  %%CITATION = ARXIV:1611.00728;%%

\bibitem{INTEL:SCIF}
``Symmetric Communications Interface (SCIF) For \Intel \XeonPhi Product Family Users Guide'',
Intel Corporation, \href{http://registrationcenter-download.intel.com/akdlm/irc_nas/9669/scif_userguide.pdf}
{http://registrationcenter-download.intel.com/akdlm/irc\_nas/9669/scif\_userguide.pdf}.


\bibitem{FBICGSTAB}
H.~Tadano and T.~Sakurai, LSSC'07, Lec. Notes Com-put. Sci. 4818, 721 (2008);
  S.~Aoki {\it et al.} [PACS-CS Collaboration],
  %``2+1 Flavor Lattice QCD toward the Physical Point,''
  Phys.\ Rev.\ D {\bf 79} (2009) 034503
  doi:10.1103/PhysRevD.79.034503
  [arXiv:0807.1661 [hep-lat]].
  %%CITATION = doi:10.1103/PhysRevD.79.034503;%%
  %412 citations counted in INSPIRE as of 13 Nov 2016

%\cite{Osaki:2010vj}
\bibitem{Osaki:2010vj}
  Y.~Osaki and K.~I.~Ishikawa,
  %``Domain Decomposition method on GPU cluster,''
  PoS LATTICE {\bf 2010} (2010) 036
  [arXiv:1011.3318 [hep-lat]].
  %%CITATION = ARXIV:1011.3318;%%
  %3 citations counted in INSPIRE as of 13 Nov 2016

\bibitem{KNL:OFP}
Joint Center for Advanced High Performance Computing (JCAHPC),
\href{http://jcahpc.jp/eng/index.html}{http://jcahpc.jp/eng/index.html}.

\bibitem{ICCPCCS}
\Intel Parallel Computing Center at Center for Computational Sciences (CCS), University of Tsukuba,
Intel Corporation,
\href{https://software.intel.com/en-us/articles/intel-parallel-computing-center-at-center-for-computational-sciences-university-of-tsukuba}
{https://software.intel.com/en-us/articles/intel-parallel-computing-center-at-center-for-computational-sciences-university-of-tsukuba}.

\end{thebibliography}
\end{document}